\documentclass[runningheads]{llncs}
\usepackage[margin=1.25in]{geometry}
\usepackage{cite}
\usepackage{stfloats}
\usepackage{amsmath,amssymb,amsfonts}
\usepackage{algorithm,algorithmicx}
\usepackage[noend]{algpseudocode}
\usepackage{graphicx}
\usepackage{textcomp}
\usepackage{hyperref}
\usepackage{multirow}
\usepackage[normalem]{ulem}

\usepackage{mathtools}
\usepackage{graphicx}
\newcommand{\norm}[1]{\left\lVert#1\right\rVert}
\usepackage{mathtools}
\usepackage{bm}
\usepackage{booktabs}
\begin{document}

\title{Joint Optimization of Hadamard Sensing and Reconstruction in Compressed Sensing Fluorescence Microscopy}
\titlerunning{End-to-end Sensing and Reconstruction in CS-FM}

\author{Alan Q. Wang*\inst{1} \and 
Aaron K. LaViolette* \inst{2} \and 
Leo Moon \inst{2} \and 
Chris Xu \inst{2} \and 
Mert R. Sabuncu\inst{1}} 

\authorrunning{A. Wang et al.}

\institute{School of Electrical and Computer Engineering, Cornell University \and School of Applied and Engineering Physics, Cornell University}

\maketitle              
\begin{abstract}
Compressed sensing fluorescence microscopy (CS-FM) proposes a scheme whereby less measurements are collected during sensing and reconstruction is performed to recover the image. 
Much work has gone into optimizing the sensing and reconstruction portions separately. 
We propose a method of jointly optimizing both sensing and reconstruction end-to-end under a total measurement constraint, enabling learning of the optimal sensing scheme concurrently with the parameters of a neural network-based reconstruction network. 
We train our model on a rich dataset of confocal, two-photon, and wide-field microscopy images comprising of a variety of biological samples. We show that our method outperforms several baseline sensing schemes and a regularized regression reconstruction algorithm. Our code is publicly-available at \url{https://github.com/alanqrwang/csfm}.
\end{abstract}
\section{Introduction}
\label{sec:introduction}
Fluorescence microscopy (FM) is an imaging modality widely used in biological research \cite{lichtman2005fluorescence}. FM follows a Poissonian-Gaussian noise model which determines the signal-to-noise ratio (SNR)~\cite{foi2008practical,zhang2019poissongaussian}. One common solution to increase SNR involves taking multiple measurements and averaging them; however, this invariably leads to an increase in the total number of photons used to generate an image, which can be undesirable when factors like photo-damage, photo-bleaching, and viability of the sample are considered~\cite{hopt2001highlynonlinear,sluder2013photodamage}. Therefore, the trade-off between image quality and the total number of measurements collected is important to optimize~\cite{sehyung2020munet,wang2018crossmodality,weigert2018care}.

Compressed sensing (CS) seeks to address this by collecting fewer measurements during sensing~\cite{gibson2020singlepixel,parot2019opticalsectioned,studer2012hyperspectral,wijesinghe2019singlepixel} and performing reconstruction on the noisy measurements to recover the image. Typically, reconstruction involves iterative optimization \cite{beck2009fista,boyd2011admm} or data-driven techniques \cite{xue2019celldetection,yao2019netflics}. 

Recently, researchers in multiple domains have explored the idea of jointly optimizing both the sensing and reconstruction parts end-to-end on training data, thus allowing for improved performance compared to separate optimization~\cite{bahadir2020loupe,sun2020computational,xue2019celldetection}. 
In this paper, we build on this idea and propose a method of jointly optimizing sensing and reconstruction in CS-FM.

Our contributions are as follows. We design a stochastic sensing model which is capable of learning the relative importance of each coefficient in the Hadamard sensing basis. Simultaneously, we train the parameters of a neural network which reconstructs the noisy measurements from the sensing model. Both parts are optimized end-to-end on training data, subject to a loss which maximizes reconstruction quality under a total measurement constraint. We evaluate both our sensing and reconstruction models and show a performance gain over CS-FM benchmarks.

\section{Background}
\subsection{Fluorescence Microscopy and Hadamard Sensing}
Let $\bm{x} \in \mathbb{R}^N$ be a vector of $N$ pixels in the field-of-view (FOV), where $\bm{x}_i$ is the signal from pixel $i$ for a fixed acquisition time.
In FM, a popular basis for collecting measurements of $\bm{x}$ is the Hadamard basis, which is well-studied, computationally-efficient, and easily realized via a digital micromirror device (DMD) in the imaging setup~\cite{streeter2009opticalhadamard,studer2012hyperspectral,sun2017russiandolls,yu2020reordering}. 

In Hadamard sensing, measurements correspond to the Hadamard coefficients. Mathematically, these coefficients are given by a linear transformation of $\bm{x}$ by a Hadamard matrix $H \in \{-1, 1\}^{N \times N}$, where each row $H_i$ is a Hadamard pattern.
To physically measure each coefficient, $H$ is reformulated as the difference between two binary, complementary matrices $H^+, H^- \in \{0, 1\}^{N \times N}$:
\begin{equation}
    H = \frac{1}{2}\left(\bm{J}+H\right) - \frac{1}{2}\left(\bm{J}-H\right) := H^+ - H^-,
\end{equation}
where $\bm{J}$ is a matrix of ones.
To obtain the measurement associated with the $i$th Hadamard coefficient, two sequential acquisitions are made by illuminating the sample according to the two binary patterns corresponding to rows $H_i^+$ and $H_i^-$. All the light for each acquisition is collected at one detector, and the Hadamard coefficient is computed by subtracting the two measurements\footnote{Note we are not considering the effects of a point-spread function here.}.

For noise considerations, we adopt a Poissonian-Gaussian noise model \cite{zhang2019poissongaussian,foi2008practical,makitalo2013optimalinversion}. Incorporating this noise model with the above complementary measurement scheme, each Hadamard coefficient $\bm{y}_i$ follows a distribution
\begin{align}
\begin{split}
    \label{eq:sensing}
    \bm{y}_i &\sim p(\bm{y}_i \mid \bm{x}) = \left[a\mathcal{P}\left(\frac{1}{a}H_i^+ \bm{x}\right) + \mathcal{N}(0, b)\right] - \left[a\mathcal{P}\left(\frac{1}{a} H_i^-\bm{x}\right) + \mathcal{N}(0, b)\right].
\end{split}
\end{align}
Here, $\mathcal{P}$ is a Poisson distribution which models signal-dependent uncertainty, while $\mathcal{N}$ is a normal distribution which models signal-independent uncertainty. Note that $a>0$ is a conversion parameter (e.g. one detected photon corresponds to a signal of $a$). 
All noise components are assumed to be independent.

To increase SNR, we consider the case where multiple measurements are collected and then averaged. 
In particular, for $S_i$ i.i.d. measurements\footnote{Note we are defining one measurement to be the computation of one Hadamard coefficient without averaging, despite the fact that this is a two-step process in the physical imaging setup.} of the $i$th coefficient $\bm{y}_i^{(1)}, ..., \bm{y}_i^{(S_i)} \sim p(\bm{y}_i \mid \bm{x})$, 
\begin{align}
\begin{split}
\label{eq:average}
    \bm{y}_i^{avg} &= \frac{1}{S_i} \sum_{s=1}^{S_i} \bm{y}_i^{(s)} \\
     &\sim \frac{1}{S_i}\left[a \mathcal{P}\left(\frac{S_i}{a}H_i^+ \bm{x}\right) + \mathcal{N}(0, S_ib)\right] - \frac{1}{S_i}\left[a\mathcal{P}\left(\frac{S_i}{a} H_i^-\bm{x}\right) + \mathcal{N}(0, S_ib)\right].
\end{split}
\end{align}
It follows that while the mean is invariant to $S_i$, the variance decreases as $S_i$ increases: 
\begin{align}
        \mathbb{E}[\bm{y}_i^{avg}] &= H_i^+ \bm{x} - H_i^- \bm{x} = H_i\bm{x}, \\
        Var(\bm{y}_i^{avg}) &= \frac{1}{S_i}\left[aH_i^+ \bm{x} + b\right] - \frac{1}{S_i}\left[a H_i^- \bm{x} + b\right].
\end{align}
Note that the number of photons coming from the sample to produce the image is proportional to the sum of the $S_i$'s, i.e. the total number of measurements. 
In this work, we pose the problem of optimizing the SNR-photon tradeoff as one of maximizing image quality subject to a constraint on the total measurements.

\subsection{Sensing and Reconstruction Optimization}
Prior work has explored the idea of jointly optimizing the sensing and reconstruction portions of computational imaging in different domains. The underlying intuition is that optimizing both parts concurrently leads to better performance than optimizing separately. Deterministic sensing patterns have been jointly optimized with neural-network based reconstruction schemes in designing the color multiplexing pattern of camera sensors \cite{chakrabarti2016multiplexing}, the LED illumination pattern in Fourier ptychography \cite{kellman2019ptychographic}, the optical parameters of a camera lens \cite{sitzmann2018achromatic}, and microscopy cell detection~\cite{xue2019celldetection}. More recently, stochastic sampling of sensing patterns have been explored in the context of CS magnetic resonance imaging \cite{bahadir2020loupe,zhang2020extending} and very-long-baseline-interferometry (VLBI) array design \cite{sun2020computational}. In this paper, we extend the idea of joint optimization of sensing strategies and neural-network based reconstruction to CS-FM.

\section{Proposed Method}
\begin{figure}[t]
\centering
\includegraphics[width=0.9\textwidth]{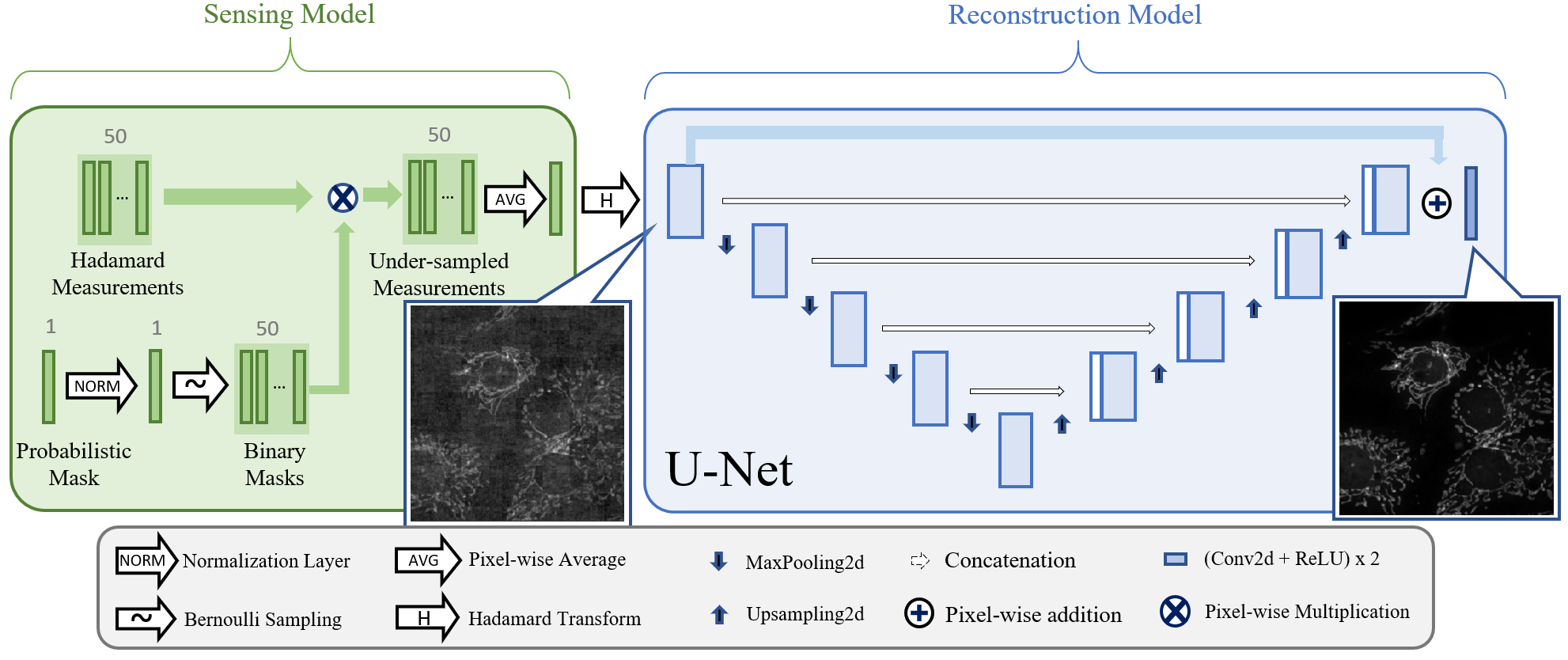}
\caption{Proposed model. Both sensing and reconstruction are trained jointly and end-to-end according to Eq.~\eqref{eq:loss}.} 
\label{fig:arch}
\end{figure}
Suppose $S$ measurements are made at each of the $N$ Hadamard coefficients. In the unconstrained case, this corresponds to $NS$ total measurements. At each coefficient, the $S$ measurements are averaged and follow a distribution according to Eq.~\eqref{eq:average}. Given a compression ratio of $\alpha \in [0, 1]$ which restricts the total number of measurements to $\alpha N S$, the question we pose is as follows: how do we learn the optimal \textit{allocation} of these $\alpha N S$ measurements across Hadamard bases, while simultaneously learning the parameters of a neural network reconstruction model, such that we maximize reconstruction quality? 

\subsection{End-to-end Sensing and Reconstruction Scheme}
Let $\bm{x}$ denote a vectorized ground truth image from a dataset $\mathcal{D}$ and let \sloppy{$\bm{y}_i^{(1)},...,\bm{y}_i^{(S)}\sim p(\bm{y}_i\mid\bm{x})$} denote $S$ measurements at the $i$th Hadamard coefficient.
Define $\bm{P} \in [0, 1]^N$ as a \textit{probabilistic mask}, normalized such that $\frac{1}{N}\norm{\bm{P}}_1 = \alpha$. 

For each Hadamard coefficient, we draw $S$ i.i.d. samples from a Bernoulli distribution with parameter $\bm{P}_i$, i.e. $\bm{M}_i^{(1)}, ..., \bm{M}_i^{(S)} \sim \mathcal{B}(\bm{P}_i)$.
In particular, $\bm{M}_i^{(s)} \in \{0,1\}$ indicates if measurement $s\in \{1,...,S\}$ should be included in the average. 
As such, we multiply $\bm{M}_i^{(s)}$ with measurement $\bm{y}_i^{(s)}$. Finally, we take the average over all included measurements to arrive at the per-coefficient measurement $\bm{z}_i$:
\begin{equation}
    \bm{z}_i = \frac{\sum_{s=1}^{S} \bm{M}_i^{(s)} \bm{y}_i^{(s)}}{\sum_{t=1}^S \bm{M}_i^{(t)}}, \ \ \text{where} \ \bm{M}_i^{(s)} \sim \mathcal{B}(\bm{P}_i) \ \text{and} \  \bm{y}_i^{(s)} \sim p(\bm{y}_i\mid \bm{x}).
\end{equation}
 If we collect all $\bm{z}_i$'s in a vector $\bm{z}$ and collect all $\bm{M}_i^{(s)}$'s in $\bm{M}$, then we can summarize the sensing model by a sensing function $\bm{z} = g(\bm{M}, \bm{x})$. Note that $\bm{M} \sim \prod_{s=1}^S\prod_{i=1}^N\mathcal{B}(\bm{P}_i)$.

$\bm{z}$ is subsequently the input to the reconstruction model, which first transforms $\bm{z}$ to image space via the Hadamard transform and then passes the result into a neural network $f_\theta$ with parameters $\theta$. In this work, we use a U-Net architecture \cite{ronneberger2015unet}. Thus, the final reconstruction is $\hat{\bm{x}} = f_\theta (H\bm{z})$.

The proposed model is illustrated in Fig.~\ref{fig:arch}.

\subsection{Loss Function}
The end-to-end sensing and reconstruction scheme is optimized jointly, according to a loss that maximizes image quality under the normalization constraint: 
\begin{equation}
    \label{eq:loss}
    \min_{\bm{P}, \theta} \mathbb{E}_{\bm{M} \sim \prod_s\prod_i\mathcal{B}(\bm{P}_i)} \sum_{\bm{x} \in \mathcal{D}} \norm{\hat{\bm{x}} - \bm{x}}_2^2 \ \ \text{s.t.} \ \ \hat{\bm{x}} = f_\theta(Hg(\bm{M}, \bm{x})) \ \ \text{and} \ \ \frac{1}{N}\norm{\bm{P}}_1 = \alpha.
\end{equation}

A higher value of the optimized $\bm{P}^*_i$ implies that more measurements for the $i$th coefficient will be averaged, resulting in a higher SNR for that coefficient. Thus, $\bm{P}^*_i$ effectively encodes the ``relative importance" of each location in Hadamard space for reconstruction. During imaging, one would draw $S$ Bernoulli realizations with parameter $\bm{P}_i^*$ to determine the number of measurements to collect at basis $i$. Note the total required measurements will be $\alpha NS$ on average.  


\subsection{Implementation}
To allow gradient computation through the Bernoulli sampling operation, we use the straight-through Gumbel-Softmax approximation \cite{jang2017gumbelsoftmax,maddison2017concrete}, which samples from a Bernoulli distribution in the forward pass and computes gradients in the backward pass via a differentiable relaxation of the sampling procedure. A temperature hyperparameter $\tau$ controls the degree of relaxation.

To enforce the normalization constraint in Eq.~\eqref{eq:loss}, we add a closed-form differentiable normalization layer to the reconstruction network\cite{bahadir2020loupe,zhang2020extending}.
Let $\tilde{\bm{P}}$ be an unconstrained probabilistic mask, and let $\bar{p} = \frac{1}{N}\|\tilde{\bm{P}}\|_1$ be the average value of $\tilde{\bm{P}}$. 
Note that $1-\bar{p}$ is the average value of $1-\|\tilde{\bm{P}}\|_1$.
Then,
\begin{equation}
    \bm{P}= 
    \begin{cases}
   \frac{\alpha}{\bar{p}}  \tilde{\bm{P}}, &\textrm{ if } \bar{p} \geq \alpha, \\
   1 - \frac{1-\alpha}{1-\bar{p}} (1 - \tilde{\bm{P}}),  &\textrm{ otherwise.}
    \end{cases}
    \label{eq:norm}
\end{equation}
It can be shown that Eq.~\eqref{eq:norm} yields $\bm{P}\in [0,1]^N$ and $\frac{1}{N}\|\bm{P}\|_1 = \alpha$, thus enforcing the hard constraint in Eq.~\eqref{eq:loss}.

\section{Experiments}
\begin{figure}[t]
\centering
\includegraphics[width=0.9\textwidth]{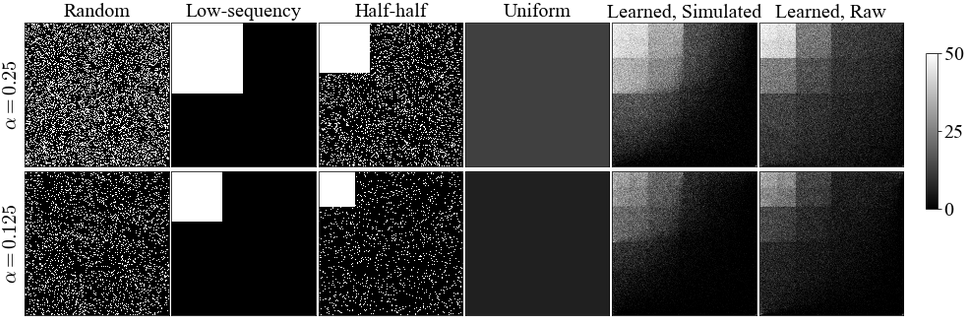}
\caption{Baseline and learned masks for $\alpha=0.25$ (top) and $\alpha=0.125$ (bottom). All masks are displayed in 2D-sequency order.}
\label{fig:masks}
\end{figure}
\subsubsection{Training Details}
In this work, we used a U-Net architecture \cite{ronneberger2015unet} for $f_\theta$, but alternative architectures which directly incorporate the forward model may be used as well, such as unrolled architectures \cite{diamond2017unrolleddeeppriors,yang2016admmnet} and physics-based neural networks \cite{kellman2019ptychographic}.
In particular, our U-Net has 64 hidden channels per encoder layer and intermediate ReLU activations. The network was trained with ADAM optimizer. We set the temperature hyperparameter $\tau=0.8$ via grid-search on validation loss. 
All training and testing experiments in this paper were performed on a machine equipped with an Intel Xeon Gold 6126 processor and an NVIDIA Titan Xp GPU. All models were implemented in Pytorch.

\subsubsection{Baselines}
For reconstruction, we compare our U-Net based method against a widely used regularized regression reconstruction algorithm (TV-W)~\cite{wang2020amortized}\footnote{We used publicly-available code from: \url{https://github.com/alanqrwang/HQSNet}.}. The algorithm minimizes a least-squares cost function with two regularization terms: total variation and L1-penalty on wavelet coefficients. The weights of these terms was established via grid search on validation performance. 

As is common in the CS-FM literature, we implemented baseline binary sensing schemes that collect either all (e.g., $S=50$) measurements or no measurement at each basis. This can be denoted with a binary mask $\bm{P} \in \{0, 1\}^N$ where $1$ indicates the basis is sampled and $0$ indicates the basis is not sampled.
Binary baseline masks include random (R), low-sequency (LS), and half-half (HH)~\cite{studer2012hyperspectral} Hadamard masks, shown in Fig.~\ref{fig:masks}. Non-binary stochastic masks include uniform (U) and learned masks. For U, on average $\alpha$-percent of all available ($S$) measurements were used at each basis, realized via Bernoulli sampling; i.e., $\bm{P}$ is a constant vector of $\alpha$'s.
We binarize the learned stochastic masks for the TV-W algorithm by taking the top $\alpha$-percentile of basis patterns. This cannot be achieved for uniform masks, so we only evaluate them on U-Net models.

\subsubsection{Data}
To illustrate the computational utility of our proposed method, models were trained and tested on a combination of in-house data and the publicly-available Fluorescence Microscopy Denoising (FMD) dataset, comprising of confocal, two-photon, and wide-field microscopy images \cite{zhang2019poissongaussian}. Our combined dataset consists of a variety of representative biological samples, namely blood vessels, neurons, and lymph nodes. Each FOV has $S=50$ noisy pixel-wise acquisitions, which consist of raw measurements at constant excitation power. Ground truth images $\bm{x}$ were obtained by pixel-wise averaging over all $S$ frames.

To obtain $\bm{y}_i^{(s)}$ input Hadamard measurements, we implemented two approaches: simulated and raw. In the simulated approach, we applied random noise to the ground truth $\bm{x}$ following Eq.~\eqref{eq:sensing}. Values of $a$ and $b$ were obtained from~\cite{zhang2019poissongaussian}\footnote{Negative values of $b$ were replaced with $0$ following the paper's experiments~\cite{zhang2019poissongaussian}.}. 
In the raw approach, we applied the Hadamard transform to raw pixel-wise measurements in a FOV. Although this is not exact since individual noise contributions are not additive, it is a reasonable approximation as we are interested in the relative importance of each basis pattern. 

All images were cropped to fixed grid sizes of $256 \times 256$.
The final training, validation, and test splits were $1000$, $100$, and $500$ cropped images, respectively; and included non-overlapping independent samples.
Reconstructions were evaluated against ground-truth images on peak signal-to-noise ratio (PSNR) and structural similarity index measure (SSIM) \cite{wang2004ssim}.

\subsection{Masks}
\begin{figure}[t]
\centering
\includegraphics[width=0.8\textwidth]{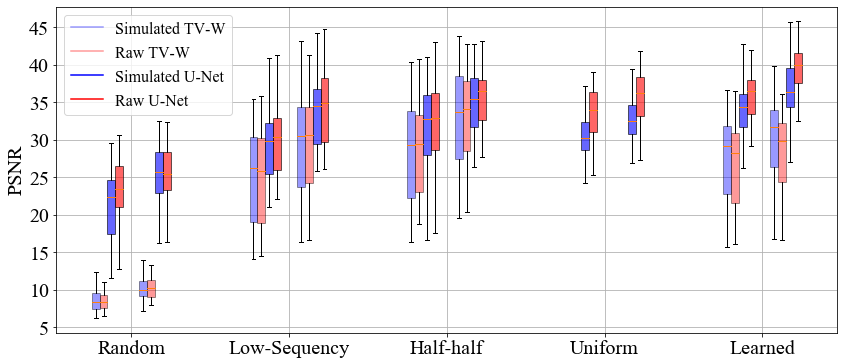}
\caption{PSNR for all masks, $\alpha=0.125$ (left) and $\alpha=0.25$ (right).} 
\label{fig:boxes}
\end{figure}
Fig.~\ref{fig:masks} shows baseline and learned masks, displayed in 2D-sequency order~\cite{gibson2020singlepixel}. We notice that the learned masks prioritize low-sequency Hadamard patterns but also spread out across higher-sequency patterns. We reason that the sensing and reconstruction scheme learns to provide the U-Net with mostly low-sequency information, and then subsequently ``assigns" the U-Net the task of imputing the missing high-sequency information.

\subsection{Reconstruction Methods}
\begin{figure}[]
\centering
\includegraphics[width=\textwidth]{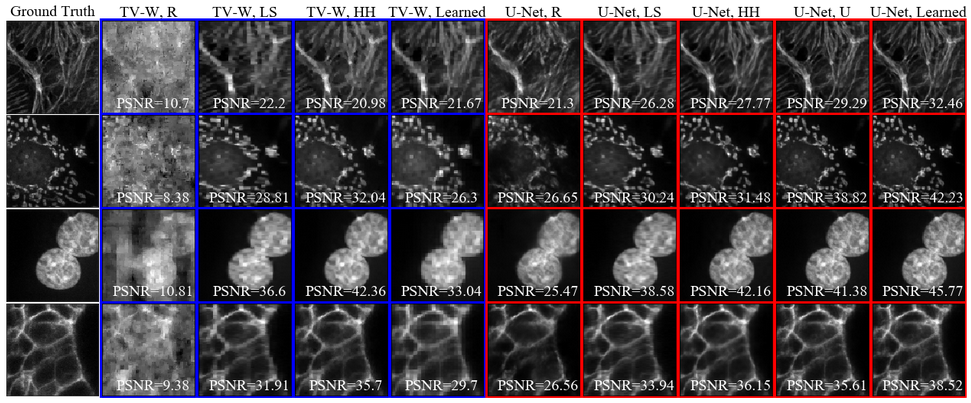}
\caption{Representative slices for $\alpha=0.25$. Left is ground truth, blue is TV-W reconstructions, red is U-Net reconstructions. Masks are indicated in titles.} 
\label{fig:slices}
\end{figure}
Fig.~\ref{fig:boxes} shows boxplots of PSNR values for both $\alpha=0.25$ and $\alpha=0.125$. U-Net reconstruction models outperform TV-W for all conditions. The pattern holds for both simulated and raw measurements.  Similarly, learned masks outperform the baseline masks consistently for U-Net reconstruction (which the sensing was optimized for), while half-half masks perform best for TV-W reconstruction. We conjecture that this is due to the U-Net model being better at imputing the missing information in the learned mask. Table 1 shows metrics for all models on PSNR and SSIM. Fig.~\ref{fig:slices} shows representative reconstructions for raw data, which allows us to qualitatively appreciate the superior quality afforded by the learned mask and U-Net reconstruction. 

\begin{table}
\label{tab}
\caption{Performance for $\alpha=0.25$ and $\alpha=0.125$. Mean $\pm$ standard deviation across test cases. Simulated/Raw.}
\begin{tabular*}{\textwidth}{@{\extracolsep{\fill}}lllcccccccc}
 $\alpha$                 & \textit{Method} & \textit{Mask} & \textit{PSNR}                            & \textit{SSIM}  \\ \toprule
 $0.25$  & TV-W & R       & 10.40$\pm$2.066/10.44$\pm$1.847          & 0.275$\pm$0.127/0.286$\pm$0.126   \\
                          &  &LS          & 29.15$\pm$6.516/29.15$\pm$6.405          & 0.887$\pm$0.094/0.886$\pm$0.095 \\
                          &  &HH          & 33.20$\pm$6.544/33.42$\pm$5.948          & 0.905$\pm$0.094/0.913$\pm$0.084 \\
                          &  &Learned     & 29.87$\pm$5.525/28.20$\pm$5.080          & 0.917$\pm$0.049/0.901$\pm$0.051 \\ \cmidrule{2-5}
                          & U-Net &R   & 25.43$\pm$3.932/25.46$\pm$3.565          & 0.692$\pm$0.124/0.692$\pm$0.135       \\
                          &  &LS       & 33.86$\pm$4.593/34.30$\pm$4.866          & 0.904$\pm$0.060/0.917$\pm$0.058          \\
                          &  &HH  & 35.04$\pm$3.882/35.80$\pm$3.942          & 0.918$\pm$0.043/0.927$\pm$0.042          \\
                          &  &U  & 31.44$\pm$6.264/35.66$\pm$3.360          & 0.841$\pm$0.125/0.924$\pm$0.024          \\
                          &  &Learned  & 36.38$\pm$4.061/39.38$\pm$3.254          & 0.932$\pm$0.044/0.953$\pm$0.018          \\
\toprule
 $0.125$ & TV-W & R      & 8.646$\pm$1.499/8.631$\pm$1.316          & 0.216$\pm$0.108/0.218$\pm$0.101   \\
                          &  &LS          & 24.91$\pm$5.984/24.90$\pm$6.042          & 0.805$\pm$0.148/0.801$\pm$0.146 \\
                          &  &HH          & 28.39$\pm$6.439/28.27$\pm$6.129          & 0.843$\pm$0.137/0.847$\pm$0.131 \\
                          &  &Learned     & 27.21$\pm$5.365/26.29$\pm$5.320          & 0.873$\pm$0.069/0.862$\pm$0.070 \\\cmidrule{2-5}
                          & U-Net &R   & 21.31$\pm$4.700/22.94$\pm$4.527          & 0.539$\pm$0.203/0.613$\pm$0.150       \\
                          &  &LS       & 29.49$\pm$4.957/30.00$\pm$4.720          & 0.813$\pm$0.129/0.826$\pm$0.125          \\
                          &  &HH  & 31.93$\pm$5.089/32.40$\pm$5.146          & 0.871$\pm$0.084/0.877$\pm$0.083          \\
                          &  &U  & 29.26$\pm$5.867/33.57$\pm$3.301          & 0.806$\pm$0.121/0.894$\pm$0.042          \\
                          &  &Learned  & 33.98$\pm$3.173/36.01$\pm$3.224          & 0.895$\pm$0.062/0.923$\pm$0.033          \\
\bottomrule
\end{tabular*}
\end{table}

\section{Conclusion}
We presented a novel sensing and reconstruction model for CS-FM, which is jointly trained end-to-end. We show that our scheme effectively learns the optimal sensing and reconstruction, and outperforms established baselines. Future work will include evaluating the proposed strategy on prospectively-collected data.

\section{Acknowledgments}
This work was, in part, supported by NIH R01 grants
(R01LM012719 and R01AG053949, to MRS), the NSF NeuroNex grant (1707312, to MRS), an NSF CAREER grant
(1748377, to MRS), and an NSF NeuroNex Hub grant (DBI-1707312, to CX).
\clearpage
\bibliographystyle{splncs04}
\bibliography{bib}

\end{document}